\documentclass[runningheads]{llncs}
\usepackage[T1]{fontenc}
\usepackage{graphicx}
\usepackage{microtype}
\usepackage{csquotes}
\newcommand{\spare}{\textsc{spare}\xspace}
\usepackage[binary-units=true]{siunitx}
\usepackage[hyphens]{url}
\usepackage[subtle]{savetrees}
\usepackage{subcaption}
\usepackage{subcaption}
\usepackage{xspace}

\usepackage{draftwatermark}
\SetWatermarkText{Preprint}
\SetWatermarkScale{4}

\begin{document}
\title{WebAssembly and Unikernels: A Comparative Study for Serverless at the Edge}
\titlerunning{WebAssembly and Unikernels: A Comparative Study}

\author{
Enrico Fiasco\inst{1}\orcidID{0009-0000-1279-8353} \and
Valerio Besozzi\inst{1,2}\orcidID{0009-0002-8493-2122} \and
Patrizio Dazzi\inst{1}\orcidID{0000-0001-8504-1503} \and
Marco Danelutto\inst{1}\orcidID{0000-0002-7433-376X}
}
\authorrunning{E. Fiasco et al.}
\institute{Department of Computer Science, University of Pisa, Pisa, 56127, Italy \and
ISTI - National Research Council, Pisa, 56127, Italy\\
\email{e.fiasco@studenti.unipi.it, valerio.besozzi@phd.unipi.it, patrizio.dazzi@unipi.it, marco.danelutto@unipi.it}}

\maketitle              % typeset the header of the contribution
\begin{abstract}
Serverless computing at the edge requires lightweight execution environments to minimize cold start latency, especially in Urgent Edge Computing (UEC). This paper compares WebAssembly and unikernel-based MicroVMs for serverless workloads. We present \textit{Limes}, a WebAssembly runtime built on Wasmtime, and evaluate it against the Firecracker-based environment used in \spare. Results show that WebAssembly offers lower cold start times for lightweight functions but suffers with complex workloads, while Firecracker provides higher, but stable, cold starts and better execution performance, particularly for I/O-heavy tasks.
\end{abstract}
\section{Introduction}
Lightweight sandboxing technologies, such as WebAssembly and unikernel-based MicroVMs, have recently gained popularity as execution environments for untrusted code in serverless platforms, particularly at the edge~\cite{10749739,gackstatter_pushing_2022}, and show promise for applications in urgent edge computing scenarios~\cite{10974814}. Compared to traditional containers, they significantly reduce cold start latency but often suffer from limited tooling, weaker platform support, and less predictable performance due to their relative immaturity~\cite{hall_execution_2019,kjorveziroski_webassembly_2023}.
Nonetheless, recent literature has demonstrated these solutions, when delineated to specific case scenarios and complemented with custom runtimes~\cite{gadepalli_challenges_2019,10.1145/3423211.3425680} and lightweight VMMs~\cite{246288,helen}, may overcome such limitations, thereby reinforcing their viability in such contexts.

In this paper, we conduct a performance comparison between WebAssembly and unikernel-based MicroVMs as lightweight sandboxing solutions for the execution of serverless functions at the edge. To achieve this goal, we present \textit{Limes}, a WebAssembly-based execution environment built on top of \textit{Wasmtime}, currently under development for \spare~\cite{10974814}. The latter is an ongoing project aimed at building a serverless platform designed for urgent edge computing scenarios.

\section{Background}
\label{sec:back}
\subsubsection{Serverless Computing}
It is a cloud execution model that enables the deployement of granularly billed and automatically scaled applications, without requiring users to manage the underlying infrastructure. 
The core concept of this paradigm are serverless functions, which are instantiated within an \textit{execution environment}, or \textit{function instance}~\cite{10.1145/3579643}. The latter typically based on classical \textit{Infrastructure-as-a-Service (IaaS)} technologies like containers or virtual machines, and deployed on demand. Thanks to the latter, this model supports fine-grained billing and horizontal scaling but introduces the well-known \textit{cold start} issue~\cite{10.1145/3587249,9191377,EBRAHIMI2024103115}. Cold start delay is caused due to serverless's characteristic to \textit{scale to zero}, as inactive function instances are deallocated, causing startup delays on subsequent requests. While cold starts cannot be fully eliminated without compromising the ability to scale to zero, they can be mitigated through pre-warming, heuristics, or optimizations at the execution environment level~\cite{10.1145/3579643}.

\subsubsection{Urgent Edge Computing}
Introduced by Dazzi et al.~\cite{10.1145/3659994.3660315}, Urgent Edge Computing (UEC) is a novel and emerging paradigm that seeks to overcome the limitations of traditional Urgent Computing by adopting a dynamic, decentralized, and highly responsive approach. UEC emphasizes processing data and making decisions close to the devices and sensors located in the affected area, aiming to reduce latency and increase the rapidity of response. The latter is an essential requirement for disaster management and emergency response scenarios.
In this context, the inherent service abstraction and dynamicity of the serverless execution model align well with the demands of UEC. In particular, the per-function execution model offered by serverless platforms enables fine-grained and adaptive resource allocation, allowing computational capacity to be provisioned and released dynamically in response to shifting demands during emergency operations. 
Moreover, a promising direction that complements this vision is the \textit{serverless edge computing} paradigm~\cite{10122638}. The latter enables the deployment of serverless applications at the edge of the network, bringing computation closer to end-users and improving resource efficiency in distributed environments. However, edge infrastructures are typically heterogeneous and often composed of resource-constrained devices, which limits the applicability of traditional virtualization and container-based solutions.
In this setting, an important aspect is the \textit{cold start latency}, which is particularly critical for near real-time applications commonly found in serverless edge and UEC scenarios (e.g., autonomous vehicles, disaster response, etc.). Since pre-warming techniques are often unfeasible in unpredictable emergency contexts, cold start can be partially mitigated through the use of lightweight execution environments such as microVMs, unikernels, and WebAssembly.

\subsubsection{MicroVMs, Unikernels and WebAssembly}
Although most serverless platforms rely on containers for deploying function instances, recent efforts in both academia and industry have explored alternative, more lightweight execution environments.
MicroVMs represent a class of lightweight virtual machines designed specifically for microservices and serverless applications. Firecracker, introduced by Agache et al.~\cite{246288}, is a KVM-based VMM optimized for fast startup and reduced overhead, and serves as the foundation for AWS Lambda and Fargate. Other notable examples include Kata Containers~\cite{8939164} and Cloud-Hypervisor~\cite{cloud_hypervisor}, both built on top of \texttt{rust-vmm}. These specialized VMMs reduce startup time and improve isolation, making them well-suited for serverless workloads.
Unikernels~\cite{10.1145/2557963.2566628}, in the other hand, represent a promising alternative to traditional containerization.
They are single-address-space machine images that offer fast boot time and provide strong isolation. 
They enables the packaging of applications by compiling the application code together with the minimal OS components required to run. Examples include MirageOS, Nanos~\cite{NanoVMs}, and OSv~\cite{10.5555/2643634.2643642}. Due to their minimal footprint and fast boot time, unikernels are increasingly considered for serverless functions deployment~\cite{10749739,helen}.
Finally, WebAssembly~\cite{WebAssemblyCoreSpecification1} has emerged as a portable binary format suitable for serverless platforms. Initially designed for web applications, it currently supports general-purpose computing thanks to runtimes like Wasmtime, Wasmer and WasmEdge. Moreover, with the introduction of WASI~\cite{WASI}, WebAssembly-based applications can interact with the underlying operating system via syscall-like interfaces, making it viable as a sandboxed and language-agnostic runtime for serverless functions~\cite{gackstatter_pushing_2022}.

\section{Prototype Design}
\label{sec:method}
\textbf{Limes}~\footnote{The source code is available at \url{https://github.com/ViktorShell/Limes/tree/vhpc}} is a lightweight asynchronous runtime manager for Wasm applications. It is written in Rust and built on top of Wasmtime~\cite{wasmtime-docs}. The latter was chosen for two key reasons. Firstly, as demonstrated in recent literature~\cite{gackstatter_pushing_2022,10.1145/3485513}, the runtime has reached a sufficient level of maturity, exhibiting consistent performance and stability. Secondly, it is known for its extensive feature set and for being WASI compliant.
Limes enables the execution of serverless functions, compiled into the \textit{wasm} binary format, within a secure multi-tenant environment. This isolation is achieved through the sandboxing mechanisms of the Wasmtime library and the inherent safety of the WebAssembly linear memory model~\cite{webasssembly-docs-security}.

\begin{figure}[ht]
    \centering
    \begin{subfigure}[b]{0.495\textwidth}
        \centering
        \includegraphics[width=0.99\linewidth]{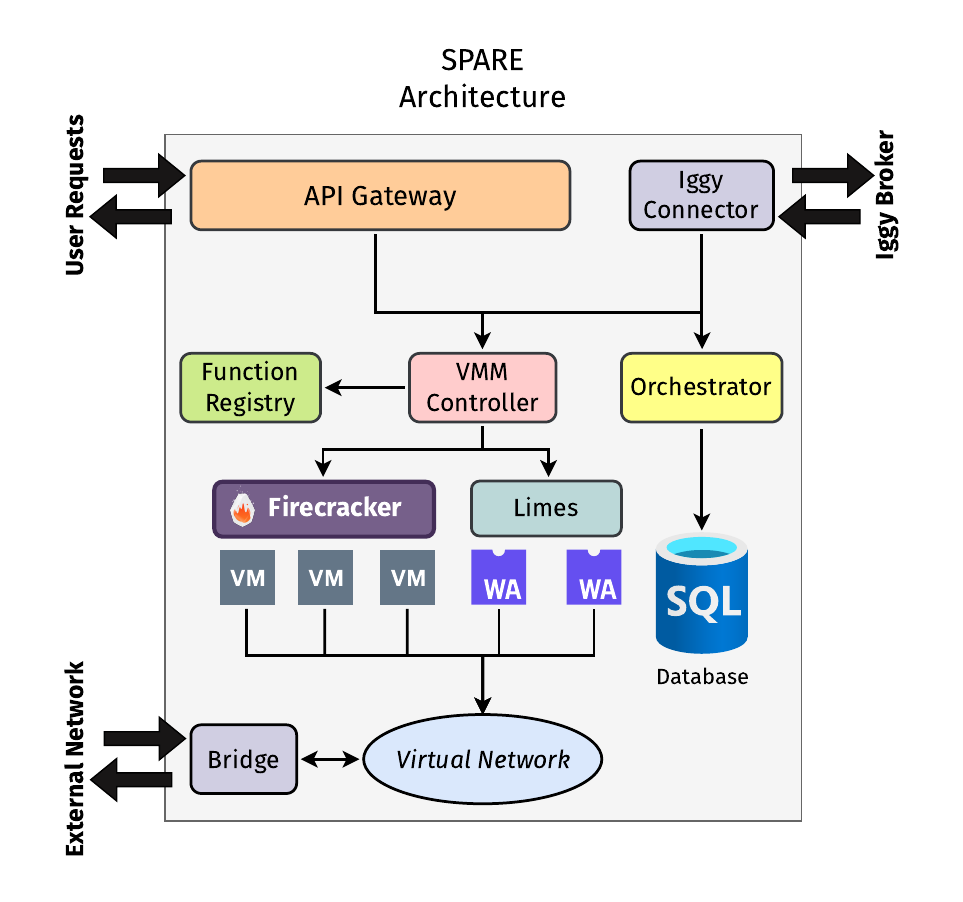}
        \caption{SPARE architecture.}
        \label{fig:spare}
    \end{subfigure}
    \hfill
    \begin{subfigure}[b]{0.495\linewidth}
        \centering
        \includegraphics[width=0.99\textwidth]{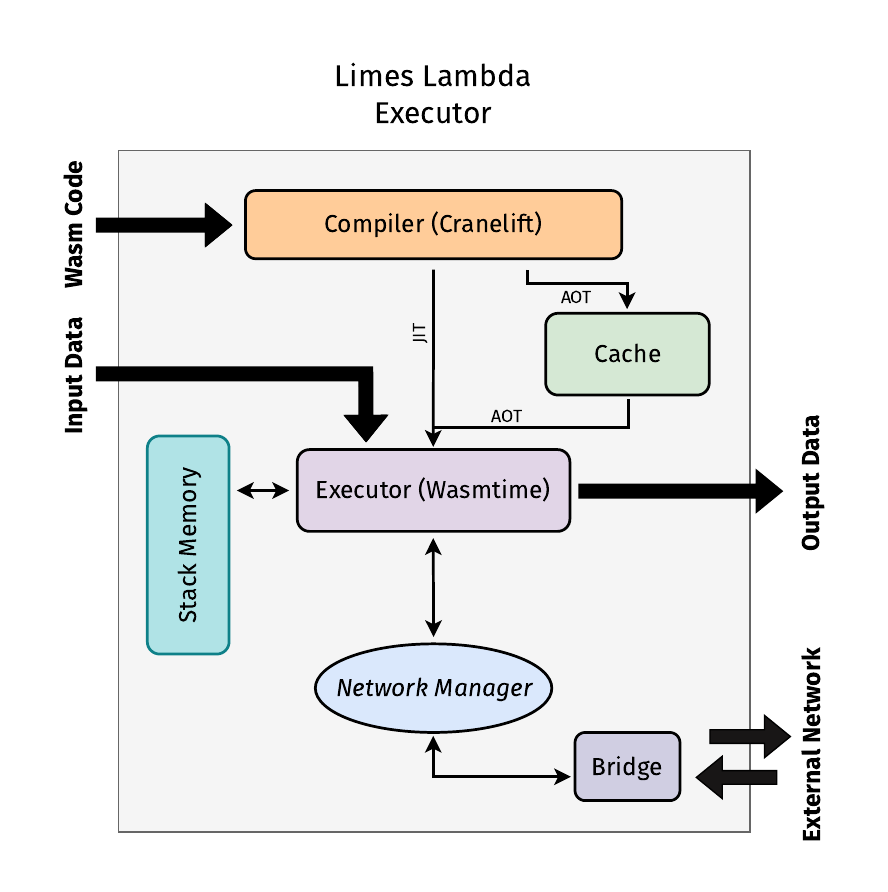}
        \caption{Limes lambda executor.}
        \label{fig:limes}
    \end{subfigure}
    \caption{Overview of the envisioned architecture.}
    \label{fig:arch}
\end{figure}

Limes is currently under development as a WebAssembly-based execution environment for \spare. Figure~\ref{fig:arch} provides the architecture of the Limes lambda executor and its intended position within the \spare platform.

\subsubsection{Architecture.}
Limes is provided in two flavors: a library implementation and a standalone binary version. The former enables its integration as execution environment in already existing code. The latter can be used independently and provides a RESTful API that allows users to register modules, initialize them, execute functions, and stop functions.
The core component of Limes architecture is its Lambda Executor component, which acts as execution environment for the deployed serverless functions. Nevertheless, depending on the flavor, Limes architecture changes slightly, with the standalone version that in addition includes a Runtime Manager, in direct contact with the API server, whose role is to handle all user requests asynchronously, and a Lambda Registry that stores the previously registered functions. Upon the invocation of a function by a user, the Runtime Manager performs a check on the Lambda Registry to verify whether the function was previously initialized, meaning that it was already compiled. If positive, the executor starts the execution; otherwise, it leverages JIT compilation. This is done using the Wasmtime serialization, storing the previously executed functions on the local database and retrieving them later, consequently reducing the initialization times by reusing the previously compiled Wasm modules. Finally, both library and standalone versions are written leveraging Rust \texttt{async} through the Tokio Runtime~\cite{tokio-runtime-docs}.

\subsubsection{Lambda Executor.}
One of the key components in Limes' architecture is the \textit{Lambda Executor}, which provides all the necessary capabilities to load a function, execute it, and return the result to the caller, effectively acting as the execution environment. The Lambda Executor revolves around the \texttt{Component} object, which abstracts a WebAssembly module, allowing it to be compiled and executed using either JIT (Just-In-Time) or AOT (Ahead-Of-Time) compilation. In this regard, Wasmtime internally leverages the Cranelift compiler, known for its fast compilation to high-quality machine code optimized for both speed and memory usage.

Currently, WebAssembly supports two execution paradigms. The first relies on wasm modules interacting directly with guest memory, while the second is based on the concept of components and uses the WebAssembly Interface Types (WIT) format~\cite{wasmtime-docs}, which defines interfaces for passing complex values like structured data and function calls between guest and host.
Limes leverages the latter approach to execute a WebAssembly module: it loads it from file and calls \texttt{Component::new}, utilizing JIT compilation~\cite{wasmtime-docs}, and pass it to the Lambda Executor. Moreover, once compiled, Wasmtime supports reusing generated machine code via serialization and deserialization mechanisms. This allows Limes to bypass the compilation pipeline in subsequent executions, thereby reducing initialization latency.

Another important feature of the Lambda Executor is its ability to interrupt the execution of a running function. Wasmtime supports two mechanisms to achieve this: the \textit{fuel} mechanism and the \textit{epoch interrupt}. The fuel mechanism offers precise interruption control at the cost of higher runtime overhead. In contrast, the epoch interrupt inspects a variable after a fixed number of instruction cycles. If an interruption is signaled, a closure is invoked, transferring control back to the caller. Limes adopts the epoch-based approach to implement function interruption.

Finally, Limes leverages the \texttt{Linker} structure from the Wasmtime library to link custom host functions into the running Wasm module. This feature enables sandboxed functions to invoke host-side logic. Through this mechanism, standard WASI capabilities~\cite{WASI} are exposed, allowing interaction with host resources such as the file system and network. In that regard, Limes employs WASI Preview 2 (WASIP2), leveraging the \texttt{wasmtime-wasi} crate.

\section{Experimental Evaluation}
\label{sec:exp}
\subsection{Experimental Setup}
To assess the performance of Limes against the default execution environment in \spare, based on Firecracker, we conducted an evaluation measuring execution times and cold start latencies of both sandboxing solutions.
We implemented three small serverless applications as workloads: \textit{no-op}, a simple function that returns the input string; \textit{mandelbrot set}, a sequential implementation of the Mandelbrot set; and \textit{image processing}, a function that reads an image from disk and applies various filters.
All applications were written in Rust, and we generated both the corresponding \texttt{wasm} module for Wasmtime and the Nanos-based unikernel image for Firecracker.
For the WebAssembly versions of \textit{mandelbrot set} and \textit{image processing}, we generated two variants: one with I/O (writing to disk), and one without I/O (only reading, in the case of \textit{image processing}), to study the impact of I/O on compilation and execution.
Finally, we conducted all the experiments on a single x86/AMD-based machine equipped with an AMD Ryzen 7 5700U (8 cores @ 1.8 GHz) and 16 GB of RAM. This configuration was chosen to mimic a realistic edge deployment environment.

\subsection{Results}
\begin{figure*}[t!]
    \centering
    \begin{subfigure}[t]{0.5\textwidth}
        \centering
          \includegraphics[width=0.99\textwidth]{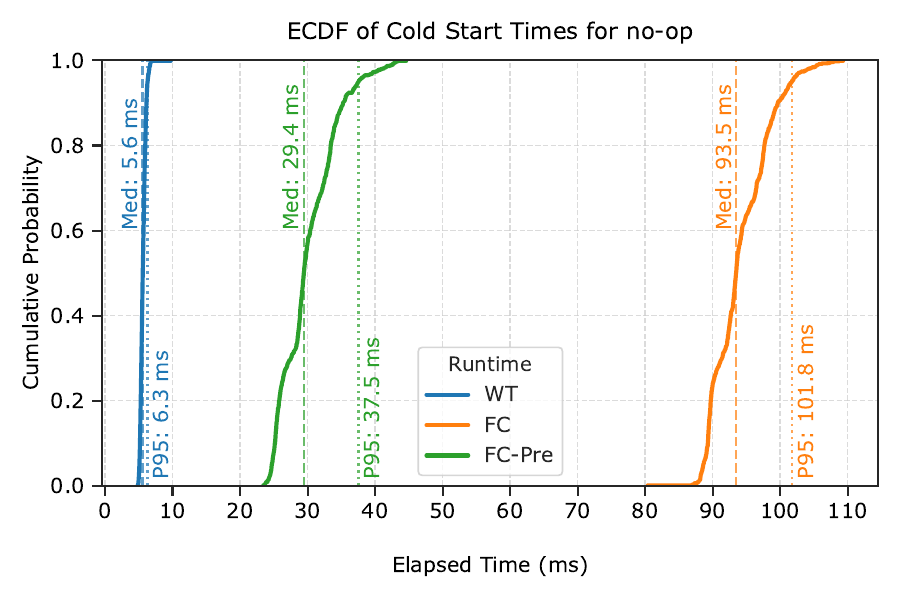}
        \caption{no-op}
        \label{subfig:no-op}
    \end{subfigure}%
    ~ 
    \begin{subfigure}[t]{0.5\textwidth}
        \centering
          \includegraphics[width=0.99\textwidth]{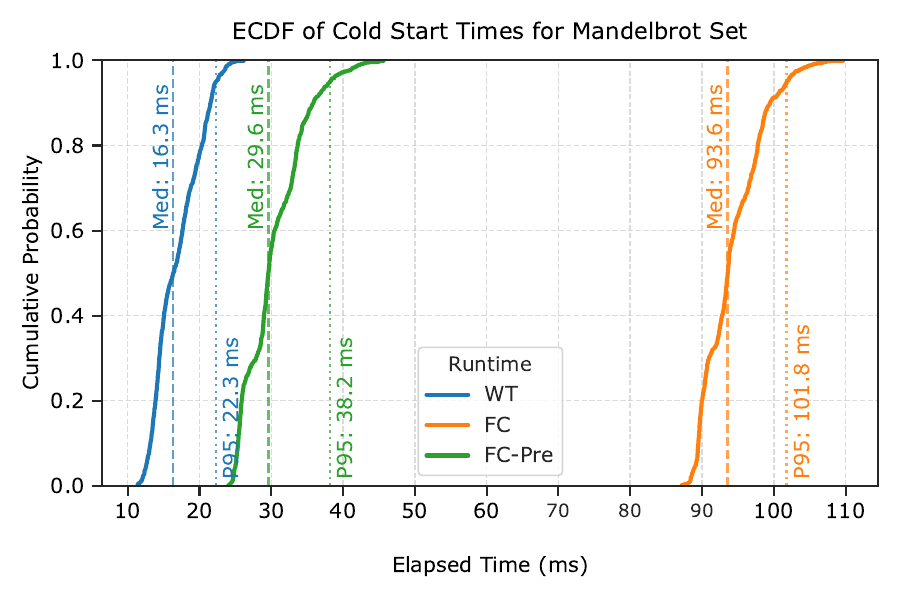}
    \caption{mandelbrot set}
    \label{subfig:mandel}

    \end{subfigure}

        \begin{subfigure}[t]{0.5\textwidth}
        \centering
          \includegraphics[width=0.99\textwidth]{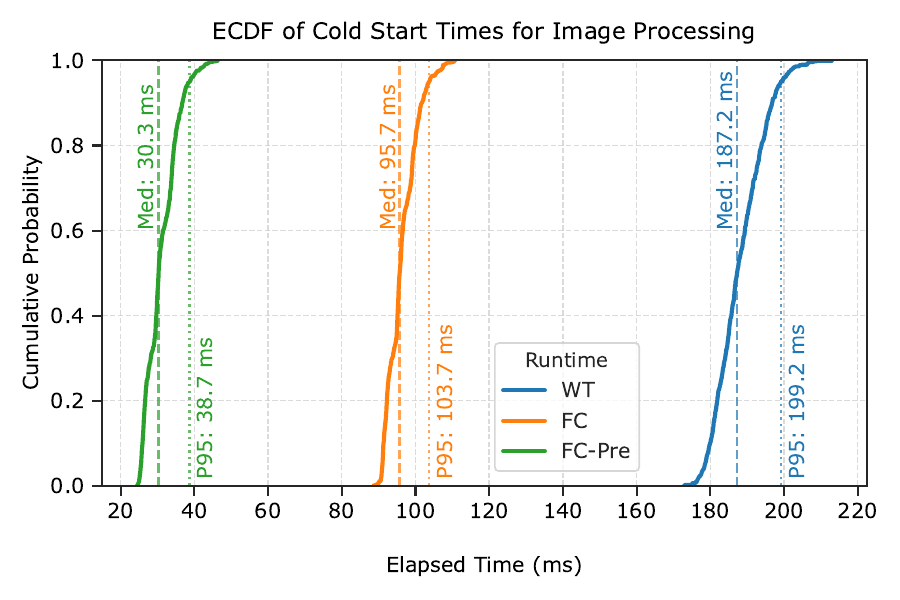}
    \caption{image-processing}
            \label{subfig:image}

    \end{subfigure}
    \caption{Empirical comulative distribution of cold start latencies for starting a function in Firecracker, pre-configured Firecracker, and Wasmtime.}
  \label{fig:cdf}
\end{figure*}

In our analysis, we measured both the execution time and the cold start latency required to instantiate each of the evaluation functions using Wasmtime and Firecracker.
For Wasmtime, we define cold start time as the interval between the invocation of \texttt{Component::new()}, which loads and compiles the WebAssembly module using JIT, and the point at which the function instance is ready to execute.
For Firecracker, we measure cold start latency as the time elapsed from the function invocation until the \spare runtime inside the MicroVM signals readiness via \textit{vsock}.
Following the methodology of Agache et al.~\cite{246288}, we report two distinct cold start measurements for Firecracker: (i) the end-to-end latency to launch an instance from scratch, and (ii) the latency to launch a VM from a pre-configured instance (i.e., with the virtual network adapter and other components already set up), considering only the final API call to the Firecracker monitor.

We conducted 1000 iterations for each configuration. Figure~\ref{fig:cdf} presents the empirical cumulative distribution function (ECDF) of cold start latencies for the three setups: \textit{wasmtime}, \textit{pre-configured Firecracker}, and \textit{end-to-end Firecracker}.
For the no-op function, whose results are shown in Figure~\ref{subfig:no-op}, Wasmtime achieved the lowest average latency at $\SI{5.6}{\milli\second}$, followed by pre-configured Firecracker with $\SI{30.1}{\milli\second}$, and end-to-end Firecracker, which recorded the highest average latency at $\SI{94.1}{\milli\second}$.
The results for the Mandelbrot set function, presented in Figure~\ref{subfig:mandel}, show a slightly increased cold start time for Wasmtime, with an average of $\SI{16.9}{\milli\second}$, followed by pre-configured Firecracker at $\SI{30.3}{\milli\second}$, and end-to-end Firecracker with $\SI{94.2}{\milli\second}$.
Finally, for what regard the image-processing function, which results are shown in Figure~\ref{subfig:image}, pre-configured Firecracker achieved the lowest cold start, with an average of $\SI{31.1}{\milli\second}$, followed by end-to-end Firecracker with $\SI{96.3}{\milli\second}$, and Wasmtime with an average of $\SI{188.0}{\milli\second}$.

\begin{figure}[t!]
    \centering
    \includegraphics[width=0.99\textwidth]{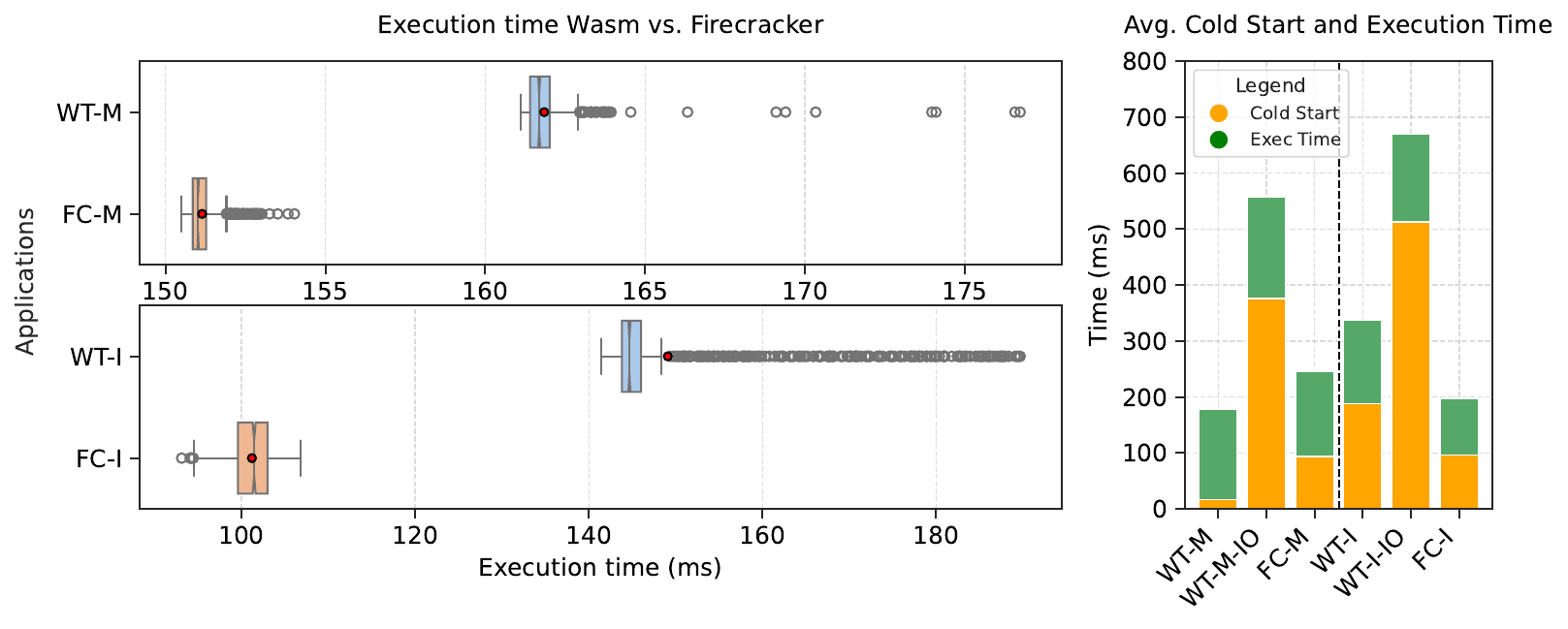}
    \caption{Average cold start and execution times of the serverless functions used for the evaluation. WT indicates Wasmtime, FC indicates Firecracker, M stands for Mandelbrot, I for Image Processing, and IO denotes whether I/O was enabled.}
    \label{fig:execution}
\end{figure}

Moreover, Figure~\ref{fig:execution} presents the measured execution times of each application considered, distinguishing between the components relative to cold start and execution. For Firecracker, the figure reports the end-to-end cold start latencies. Regarding execution times, Firecracker achieves the lowest values for both applications: \SI{151.14}{\milli\second} for the Mandelbrot set and \SI{101.20}{\milli\second} for image processing. In comparison, Wasmtime records \SI{161.84}{\milli\second} and \SI{149.09}{\milli\second} respectively for the Mandelbrot set and image processing functions without IO enabled. When executed with IO enabled, Wasmtime recorded increased execution times of $\SI{180.72}{\milli\second}$ for the Mandelbrot set and $\SI{157.86}{\milli\second}$ for image processing.

These results indicate that while Wasmtime demonstrates the lowest cold start latency for lightweight functions, it is more sensitive to the complexity of the code being instantiated. As the function logic becomes more complex, as seen in the image-processing function that relies on the external \textit{image} crate, initialization and compilation times increase significantly, leading to higher latencies.
Conversely, Firecracker exhibits more stable cold start times regardless of function complexity. This derives from the fact that much of the initialization overhead, including OS bootstrapping and runtime setup, is decoupled from the application logic itself, which is compiled AOT and packaged into a unikernel-based vm image. Nevertheless, the end-to-end startup cost of Firecracker remains higher. However, this can be substantially reduced by leveraging pre-configured microVMs, which greatly shorten the launch path.

\section{Related Work}
\label{sec:related}
Several studies have investigated alternative approaches to traditional containerization as execution environment for serverless computing, aiming to mitigate cold start latency, enhance functions portability, and improve overall performance.

Unikernels, due to their lightweight nature, have emerged as a promising solution to encapsulate serverless functions.
In this context, Mistry et al.~\cite{9407311} proposed \textit{UniFaaS}, a solution tailored for edge environments based on MirageOS and Solo5. Experiments showed reduced boot times and improved performance over Docker-based environments used in Apache OpenWhisk. Fingler et al.~\cite{10.1145/3343737.3343750} introduced \textit{USETL}, employing the Rumprun unikernel and optimizing network I/O via a high-level communication interface. Similarly, Tan et al.~\cite{9213020} developed \textit{UaaF}, achieving efficient inter-function communication using \texttt{VMFUNC} to reduce hypervisor-induced overheads.
Cadden et al.~\cite{10.1145/3342195.3392698} presented \textit{SEUSS}, which combines the Rumprun unikernel and VM snapshots for rapid function deployment and reduced cold starts.
Lightweight virtual machine monitors (e.g., Firecracker and Cloud Hypervisor) further help reduce cold start latency and integrate well with unikernels. Mavridis and Karatza~\cite{helen} evaluated various lightweight VMMs and showed how combining them with unikernels effectively reduces cold starts. Moebius et al.~\cite{10749739} extended the analysis to Nanos and OSv with Firecracker, finding them superior to Docker in cold start time and memory usage.

Recent work has also explored WebAssembly as a lightweight execution environment for serverless workloads. Long et al.~\cite{9214403} demonstrated that the use of it achieved faster execution times compared to Docker, while Kjorveziroski et al.~\cite{kjorveziroski_webassembly_2023} provided a comprehensive comparison of wasm runtimes, showing that AOT compiled WebAssembly significantly reduces cold starts and improves execution.
Hall et al.~\cite{hall_execution_2019} implemented a serverless platform prototype using Node.js with V8 as the WebAssembly runtime and evaluated its use for serverless edge computing. However, experimental evaluation showed slower execution times compared to Apache OpenWhisk, but notably reduced cold start latency. 
The mixed results in the literature suggest that the performance of WebAssembly depends primarily on the runtime used.
In this regard, Gackstatter et al.~\cite{gackstatter_pushing_2022} benchmarked and compared multiple WebAssembly runtimes to assess their performance and suitability in edge environments.
Although performance is mainly runtime-dependent, the authors highlighted the current limitations of WebAssembly due to its relatively early maturity.
Hence, a key area of research is the development of WebAssembly runtimes that are specifically tailored to particular use cases and deployment scenarios.
Addressing this, Gadepalli et al.~\cite{gadepalli_challenges_2019,10.1145/3423211.3425680} introduced \textit{aWsm} and \textit{Sledge}, a WebAssembly runtime and a WebAssembly-based serverless framework designed for edge deployment.
Sledge, built on aWsm, uses AoT compilation to LLVM IR to optimize WebAssembly-based serverless functions for low-latency execution, while enabling concurrency and distributed workloads for efficient edge deployment.

\section{Conclusions}
\label{sec:conclusion}
This paper presented a performance evaluation of WebAssembly against unikernel-based MicroVMs in the context of urgent edge computing and, more broadly, serverless at the edge. A key requirement for these scenarios is minimal \textit{cold start latency}, crucial to support real-time applications. In UEC, pre-warming techniques are often impractical due to the unpredictable nature of emergency events, making lightweight and fast execution environments essential.
To address this, we developed \textit{Limes}, a WebAssembly-based execution environment built on Wasmtime, and compared it to the one used in \spare, that leverages Firecracker and unikernel-based MicroVMs. Our results show that Wasmtime achieves lower cold start latencies, particularly for lightweight functions, but is more sensitive to function complexity, with delays increasing as dependencies or internal logic grow. In contrast, Firecracker offers stable cold start times regardless of function complexity, though with generally higher end-to-end startup times.
While Firecracker outperforms WebAssembly in execution speed overall, WebAssembly remains a viable alternative for lightweight serverless workloads in UEC scenarios. It balances fast startup times with manageable execution overheads, therefore mitigating JIT costs while delivering competitive performance.

\begin{credits}
\subsubsection{\ackname} 
This work has been partially funded by the NOUS (A catalyst for EuropeaN ClOUd Services in the era of data spaces, high-performance and edge computing) HORIZON-CL4-2023-DATA-01-02 project, G.A. n. 101135927,
and by
Spoke 1 ”FutureHPC \& BigData” of the Italian Research Center
on High-Performance Computing, Big Data and Quantum Computing (ICSC)
funded by MUR Missione 4 Componente 2 Investimento 1.4: Potenziamento
strutture di ricerca e creazione di ”campioni nazionali di R\&S (M4C2-19 )”
- Next Generation EU (NGEU).
\end{credits}
\bibliographystyle{splncs04}
\bibliography{bib}
\end{document}